\documentstyle[12pt,fleqn]{article}

\newcommand{\bc}{\begin{center}}
\newcommand{\ec}{\end{center}}

\newcommand{\be}{\begin{equation}}
\newcommand{\ee}{\end{equation}}

\newcommand{\Qox}{\begin{picture}(6,6)(150,200)
                    \put(150,200){\framebox(6,6)}
                  \end{picture}}

\newcommand{\LR}{\small {\hspace*{-0.8cm}
\begin{array}{c}
     L\vspace*{-0.5cm}\\ \,\,\,\,\,\,/ \vspace*{-0.3cm}\\
        \,\,\,\,\,\,\,\,\,\,R
  \end{array} \hspace*{-0.5cm}}}
\newcommand{\RL}{\small {\hspace*{-0.8cm}
\begin{array}{c}
     R\vspace*{-0.5cm}\\ \,\,\,\,\,\,/ \vspace*{-0.3cm}\\
        \,\,\,\,\,\,\,\,\,\,L
  \end{array} \hspace*{-0.5cm}}}
\begin{document}
\renewcommand{\theequation}{\thesection.\arabic{equation}}
\renewcommand{\thefootnote}{\fnsymbol{footnote}}
\title{New Higgs Field Ansatz for Effective Gravity in Flat Space Time}
\author{Andreas Geitner\footnotemark, Matthias Hanauske\footnotemark, 
Eckhard Hitzer\footnotemark}
\date{}
\maketitle
\addtocounter{footnote}{1}
\footnotetext{geitner@spock.physik.uni-konstanz.de}
\addtocounter{footnote}{1}
\footnotetext{06994415680-0001@t-online.de}
\addtocounter{footnote}{1}
\footnotetext{hitzer@kurims.kyoto-u.ac.jp}
\begin{abstract}
Regarding Pauli's matrices as proper Higgs fields one can deduce an 
{\em effective} approximation for gravity in flat space. In this work we 
extend this approximation up to the second order. Reaching complete 
agreement in the special case of gravitational waves. Unification  
in view, we
introduce isospinorial degrees of freedom. In this way the mass spectrum 
and chiral asymmetry can be generated with the help of an addtional  
scalar
Higgs field. The Higgs modes corresponding to gravity are discussed. 
\end{abstract}
\renewcommand{\thefootnote}{\arabic{footnote}}
\setcounter{footnote}{0}
\section{Introduction}
In a sequence of previous works \cite{prev} a semiclassical theory for 
effective
gravity in flat space time was proposed. Key ingredients were:
(a) a new $SU(2) \times U(1)$ gauge of the handed chirally represented 
particle/antiparticle doublett,
(b) replacing the usual scalar Higgs field
of the standard model by assigning the Pauli matrix fields\footnote
{Already spacetime dependent because of (a).}
$\tilde{\sigma}(x)$ the role of a nonscalar Higgs field.
By analyzing the resulting new set of field equations one finds that
(c) the excitations of this new Higgs field effectively
describe Einstein's gravity in lowest order,
(d) passive gravitational and inertial mass of particles and  
antiparticles
alike are generically identical, and
(e) the gravitational constant $G$ is defined through choosing the ground
state.

These results need to be extended in primarily two directions:
(1) going from the lowest linear order to the by now experimentally
verified higher (2nd) order and
(2) by taking the other fundamental interactions into account.
New works in these two areas will be presented in the subsequent
two sections.\footnote{See \cite{prev2} for previous works in  
direction (2).}
%
\section{The Second Order\protect\footnotemark}\footnotetext{
This section is based on \cite{Han}.}
The expression for the kinetic energy of the new nonscalar
Higgs field $\tilde{\sigma}(x)$
\be
tr[(\eta^{\rho\beta}{}\eta^{\mu\nu}
                                D_\rho \tilde{\sigma}_{\mu R})]
(D_\beta \tilde{\sigma}_{\nu L}) + (\mbox{index permutations})
\label{e1}
\ee
in \cite{prev} is quadratic in $\tilde{\sigma}(x)$ ($D_\rho$ is the new
$SU(2) \times U(1)$ gauge covariant derivative and
$\stackrel{\mbox{\hspace*{-0.2cm}}(0)}
{\tilde{\sigma}^\mu}_{R/L}=\frac{v}{4}\,(1,\pm \sigma_i), v=const.$ 
defines the groundstate). The resulting field equations are
therefore only linear in $\tilde{\sigma}$, resp. its excitations
$\epsilon^\mu{}_\nu$.
($\tilde{\sigma}_\mu=(\delta^\mu{}_\nu+\epsilon^\mu{}_\nu)
\stackrel{\mbox{\hspace*{-0.2cm}}(0)}
{\tilde{\sigma}^\nu}$). Nonlinear effects like as gravitational self 
interaction, perihel shift, gravitational waves, etc. require to
generalize the Lagrange density so as to include 2nd order effects.

\subsection{Generalization}
The principle applied here is to replace in (\ref{e1}):
\be
\eta^{\rho\beta} \rightarrow\, a \,\,
tr(\tilde{\sigma}^\mu_R{}\tilde{\sigma}^\nu_L)\, , \,\,\ldots
\ee
where $a$ is at this stage just a suitably to be determined constant.
In this way the number of terms in (\ref{e1}) increases greatly and it
proved suitable to utilize the package Mathtensor of Mathematica in 
order to deduce and expand the field equations. As the comparison
with gravity in higher orders is the main concern of this section,
everything is simplified by neglecting matter- and gauge fields.
Similarly no attention was paid to the possible handedness of the
excitations $\epsilon^\mu{}_\nu$.

The expansion of the Higgs field equations
\be
{\cal H}_{\mu\nu}=\stackrel{(1)}{\cal H}_{\mu\nu}+
\stackrel{(2)}{\cal H}_{\mu\nu}+\ldots = 0\hspace{1cm} (\mbox{vacuum!})
\ee
in terms of $\epsilon^\mu{}_\nu$:
\be
\tilde{\sigma}^\mu=(\delta^\mu{}_\nu +
\stackrel{\mbox{\hspace*{-0.2cm}}(1)}{\epsilon^\mu}_\nu +
\stackrel{\mbox{\hspace*{-0.2cm}}(2)}{\epsilon^\mu}_\nu)
\stackrel{\mbox{\hspace*{-0.2cm}}(0)}{\tilde{\sigma}^\mu},
\hspace{0.5cm}
\mid{\stackrel{\mbox{\hspace*{-0.2cm}}(2)}{\epsilon^\mu}_\nu} \mid
\,\,\simeq\,\,
 \mid{ \stackrel{\mbox{\hspace*{-0.2cm}}(1)}{\epsilon^\mu}_\nu}\mid^2 
\,\,\ll\,\,
\mid {\stackrel{\mbox{\hspace*{-0.2cm}}(1)}{\epsilon^\mu}_\nu}\mid
\,\,\ll 1
\ee
yields order by order:
\be
\stackrel{(1)}{\cal H}_{\mu\nu}=0 \hspace{0.5cm}
\stackrel{\mbox{\small see$\,$\cite{prev}}}
\longrightarrow
\hspace{0.5cm}
\Qox \hspace*{-0.1cm}
\stackrel{\mbox{\hspace*{0cm}}(1)}{\epsilon}_{\mu\nu}\,=\,\,
\stackrel{\mbox{\hspace*{-0.4cm}}(1)}{\epsilon_\mu{}^\alpha}{}
_{\hspace*{-0.2cm}\mid \alpha \mid \nu}\, +
\stackrel{\mbox{\hspace*{-0.4cm}}(1)}{\epsilon_\nu{}^\alpha}{}
_{\hspace*{-0.2cm}\mid \alpha \mid \mu}
\ee
\be
\stackrel{(2)}{\cal H}_{\mu\nu}=0.
\label{e2}
\ee
As previously shown \cite{prev}
$\stackrel{\mbox{\hspace*{-0.2cm}}(1)}{\epsilon^\alpha}_\alpha
=\stackrel{(1)}{\epsilon}=0$ corresponds to 1st order energy-momentum
conservation. Expecting the same for the 2nd order should yield
$\stackrel{(2)}{\epsilon}=0$. Further analysis proved that
$\stackrel{\mbox{\hspace*{0cm}}(1)}{\epsilon}_{\alpha\beta}
\stackrel{\mbox{\hspace*{0cm}}(1)}{\epsilon}{\hspace*{-0.1cm}}^{\alpha\beta}
\,\,=0$ should hold in addition, in order to achieve some  
reasonable
comparison with gravity. Whether both of these 2nd order trace
conditions are to be explained by ways of some conservation law
remains to be investigated.

\subsection{Comparing $\stackrel{(2)}{\cal H}_{\mu\nu}=0$ with GR}
The strategy adopted is to extract an {\it effective} metric
from (\ref{e2})
\begin{eqnarray}
 g^{\mu\nu}_{\mbox{\small eff.}}  &=&
\underbrace{\eta^{\mu\nu} +  A \,\stackrel{(1)}{\epsilon}{}^{\mu\nu}}
_{\mbox{lin. case}}
+  C \,\stackrel{(2)}{\epsilon}{}^{\mu\nu}
+ D \,\stackrel{(1)}{\epsilon}{}^\mu{}_\alpha
\stackrel{(1)}{\epsilon}{}^{\alpha\nu}
\end{eqnarray}
and to express $\stackrel{(2)}{\cal H}_{\mu\nu}=0$ explicitely in
terms of it. This then allows to formally compare (\ref{e2}) with
Einstein`s equations in vacuum
\be
\stackrel{(1)}{R}_{\mu\nu}=0\hspace{0.5cm}\mbox{(lin. case \cite{prev})}
\ee
\be
\mbox{and}\stackrel{(2)}{R}_{\mu\nu}=0,
\label{e3}
\ee
where $R_{\mu\nu}$ is calculated in terms of the Christoffel symbols
formally associated with $g^{\mu\nu}_{\mbox{\small eff.}}$.

The result is an agreement of (\ref{e2}) with (\ref{e3}) up to
an antisymmetric divergence term
\be
 B_{\mu\nu}=X_{\mu\nu\sigma}{}^{|\sigma},
  \hspace*{1cm}X_{\mu\mbox{\boldmath ${\nu\sigma}$}}
=-X_{\mu\mbox{\boldmath ${\sigma\nu}$}}
\ee
\be
\hspace*{-1cm}X_{\mu\nu\sigma} =
   - v \, \epsilon\,_{[\nu}{}_\rho\,\epsilon_\mu{}^\rho{}_{|\sigma]} - 
  v \, \epsilon_\mu{}_\rho\,\epsilon\,_{[\nu}{}^\rho{}_{|\sigma]}
     - v \,  \epsilon_\rho{}^\gamma\,\eta_\mu{}_{[\nu}\,
        \epsilon_{\sigma]}{}^\rho{}_{|\gamma} -
  v \, \epsilon\,_{[\sigma}{}_\rho\,\eta_\mu{}_{\nu]}\,
       \epsilon^\rho{}^\gamma{}_{|\gamma}
\ee
The necessary restrictions on $g^{\mu\nu}_{\mbox{\small eff.}}$ are  
simply
$A\neq0$ and $D= \frac{1}{2}A^2{}-1$. $C$ remains free.
The antisymmetry of $X$ means that it plays no role in the  
energy-momentum conservation. By further restricting $A=C=2$\,\, it  
becomes possible to write
\be
g^{\mu\nu}_{\mbox{\small eff.}}{}\,\,\mbox{\bf 1}=(\frac{4}{v})^2 \,\,
        \tilde{\sigma}^{(\mu}_R\tilde{\sigma}^{\nu)}_L
\ee
Furthermore (\ref{e2}) may be rewritten as
\begin{displaymath}
\hspace*{-1cm}
\stackrel{(2)}{S}_{\mu\nu}{}^\alpha{}_{|\alpha}=
co. \left( \stackrel{(2)}{t}_{(\mu\nu)}(\tilde{\sigma})
- \frac{1}{2} \stackrel{(2)}{t}\hspace*{-0.2cm}(\tilde{\sigma}) \,  
\eta_{\mu\nu}\right);
\,\, t_\mu{}^\nu = \sum_{R,L}\frac{\partial {\cal L}_{Higgs}}
{\partial \tilde{\sigma}^\alpha{}_{|\nu}}\,\,  
\tilde{\sigma}^\alpha{}_{|\mu}
- \,\,\delta_\mu{}^\nu{} {\cal L}_{Higgs}
\end{displaymath}
if $A^2 = 2\,\, C$. This strongly resembles analogous expressions in GR 
\cite{DFC}
and shows the analogy wrt. self interaction in both theories.
In the case of gravitational waves, it finally turns our that by
applying the socalled short wave formalism \cite{MTW}
the $B_{\mu\nu}$-discrepancy
averages out to zero.
%
\section{Isospinorial Extended  
Theory\protect\footnotemark}\footnotetext{Compare
gr-qc/9712074 for further details.}
\subsection{The Lagrangian}
We start with a chirally symmetric representation in which we put
leptons and quarks in a 4-iso spinor $ \psi_{L/R}=\left(
\begin{array}{c} \nu \\ e \\ u \\ d \end{array} \right)_{L/R} $,
where each entry itself is a 2-spinor $ \nu _{L/R} = \left(
\begin{array}{c}\nu \\ \bar{\nu} \end{array}\right) $.
Most parts of the Lagrangian have now the same form as in the
isoscalar version:
$$\hspace{-2.7cm}\mbox{The fermionic term:}\hspace{1cm}
{\cal L}_M=\tilde{\psi}_L^{\dag}\tilde{\sigma}^\mu_LD_ \mu
\tilde{\psi}_L + h.c.+ (L\leftrightarrow R)
$$
The Higgs field term:
\begin{eqnarray*}
{\cal L}_{H1}=tr(D_ \alpha\tilde{\sigma}_L^\mu D^\alpha
\tilde{\sigma}_{r \mu}- D_ \alpha\tilde{\sigma}_L^\alpha D_\mu
\tilde{\sigma}_{R}^\mu- D_ \alpha\tilde{\sigma}_L^\alpha D_\mu
\tilde{\sigma}_{R}^\mu) - \\
\mu^2  tr(\tilde{\sigma}^\mu \tilde{\sigma}_ \mu)
-\frac{\lambda}{12}\ tr(\tilde{\sigma}^\mu \tilde{\sigma}_ \mu)^2
\end{eqnarray*}
$$\hspace{-2.6cm}\mbox{The Yukawa coupling term:}\hspace{1cm}
{\cal L}_I=-k(\tilde{\psi}_L ^{\dag}\tilde{\sigma}_{L
\mu}\tilde{\sigma}_R^\mu \tilde{\psi}_R + h.c.)
$$
To be able to generate all fermionic and bosonic masses we need an  
additional scalar Higgs field, who's Lagrange density is
$$
{\cal L}_{H2}=(D_ \alpha \phi)^{\dag}(D^\alpha \phi) -
\frac{\bar{\mu}}{2}\,\phi ^{\dag} \phi -
\frac{\bar{\lambda}}{4}\,(\phi ^{\dag} \phi)^2
$$
And of course we need the kinetic term for the gauge bosons, which  
has the usual form since we do not yet couple them to gravity:
$$
{\cal L}_F=-\frac{1}{16 \pi}F_{\mu \nu a}F^{\mu \nu a}
$$
This Lagrangian is invariant under
$U(1)\times SU(2)_{spin}\times SU(2)_{isospin}$ transformations  
with the
generators
$$
\tau_i=\frac{1}{2}\,1|_{spin}\left(\begin{array}{cc}\sigma_i & 0 \\  
0 & \sigma_i\end{array}\right) \: ,i =0..3 \\
\,\,\,\,\tau_i=\frac{1}{2}\,1|_{isospin}\sigma_{i-3} \: ,i=4..6
$$
were $\sigma_i$ are the usual Pauli-matrices ($\sigma_0\equiv  
\mbox{\bf 1}$).
\subsection{Spontaneous Symmetry Breaking}
First we choose a Basis: Spin space as usual
$\sigma_{\hspace{0.3cm}\RL}^\mu\,\,=(\sigma^0,\pm\sigma^i)$  
\vspace*{-0.3cm}\\
\hspace*{40pt} Isospin space $N^a=\left(\begin{array}{cc}\sigma^a &  
0 \\ 0&0\end{array}\right) \,,a=0..3 $
$N^a=\left(\begin{array}{cc}0&0\\0&\sigma^a\end{array}\right) \,,a=4..7$
\hspace*{40pt} where possible quark-lepton mixing is neglected for  
simplicity.

\noindent We now write the tensor-field as ground- and exited state:\\
\hspace*{80pt} $\tilde{\sigma}_{L/R}^\mu= \,\,
\stackrel{(0)\hfill}{\tilde{\sigma}_{L/R}^\mu}
+\,\,\varepsilon_{L/R\nu a}^{\;\mu} \sigma_{L/R}^\nu N^a $
\\
\\
\noindent The Dirac equation is best reproduced if we choose
following ground-state:
$$
\stackrel{(0)\hfill}{\tilde{\sigma}_{L/R}^\mu}=\sigma_{L/R}^\mu\stackrel{(0)}{N}_{L/R}  
\, \mbox{with}  
\stackrel{(0)}{N_{L/R}}\,\,=diag\left(n_1,n_2,n_3,n_4\right)
_{L/R}
$$
where $N_{L/R}$ cannot depend on $\mu $ because of the isotropy of space.
Now we can generate chiral asymmetry by choosing $n_{R1}=-n_{R2}$.  
As result we have a right-handed neutrino, that does not couple to  
the $ W^{\pm} $ bosons, but to the Z-Boson, which is not in  
contradiction
to experiments,
since these give only evidence to the fact, that right-handed
neutrinos do not participate in week decay (see Wu-experiment
\cite{Wu}).\\
For simplicity we now choose the following ground-state, which sets 
both quark masses to be equal:
$$
N_L =\left(
\begin{array}{ll}
l {\bf 1} & 0 \\
0 & q {\bf 1}
\end{array}
\right) ,\; N_R =
\left(
\begin{array}{ll}
-l \sigma ^3 & 0 \\
0 & q {\bf 1}
\end{array}
\right)
$$
A direct consequence of the parity violation is the fact, that the  
neutrino receives a negative mass from the tensorial Higgs field.
This can be compensated by the scalar Higgs field, whose
ground-state we choose as
$$
\stackrel{0}{\phi}= v
\left( \begin{array}{c}
1 \\
0\\
0\\
0
\end{array}
\right)
$$
\subsection{Fermionic Field Equations}
The field equations for the fermions are
$$
i \tilde{\sigma} ^\mu _{\hspace{0.2cm}\RL}\, D_ \mu \psi  
_{\hspace{0.3cm}\RL}  \,\,\, +
\frac{i}{2} (D_ \mu \tilde{\sigma} ^\mu  
_{\hspace{0.2cm}\RL}{}\,\,\, ) \psi _{\hspace{0.3cm}\RL}{\,\,\,} -
k \tilde{\sigma} ^\mu _{\hspace{0.2cm}\RL}{\,\,\,}
\tilde{\sigma} _{\mu}{}_{\hspace{0.2cm}\LR}{\,\,\,}  
\psi_{\hspace{0.2cm}\LR}
\,\,\, -
\, \tilde k \phi( \phi^{\dagger}  \psi  
_{\hspace{0.2cm}\LR}{\,\,\,\,\,} ) = 0
$$
To be able to compare this to the Dirac equation of the standard theory  
we need to renormalize the fermionic spinor components:
$$
\tilde \nu _ {L,R}=\sqrt l \nu _ {L,R}, \,\tilde e_ {L,R} =  \sqrt  
l  e_{L,R} ,\,\tilde u _ {L,R}  =  \sqrt q u _ {L,R} ,\, \tilde d _  
{L,R}  =  \sqrt q d _ {L,R}
\label{e4}
$$
For the ground state this gives
$$\begin{array}{l}
0= i \sigma^\mu_R  \partial_ \mu \left(\begin{array}{c} \tilde{\nu} \\
\tilde{e}\\ \tilde{u}\\ \tilde{d} \end{array} \right)_R
\hspace*{-0.4cm}-k\left(\begin{array}{c} l\tilde{\nu} \\
l\tilde{e}\\ q\tilde{u}\\ q\tilde{d} \end{array} \right)_{L}
\hspace*{-0.4cm}+\bar{k}\frac{v^2}{l}\left(\begin{array}{c}
\tilde{\nu} \\
0\\ 0\\ 0 \end{array} \right)_{L}
\hspace*{-0.4cm}-\frac{1}{2} \left( g_1 \omega_{\mu0}\sigma^\mu_R
\left(\begin{array}{c} \tilde{\nu} \\
\tilde{e}\\ \tilde{u}\\ \tilde{d} \end{array}
\right)_R\right.\\[0.4cm]\left.
+g_2 \omega_{\mu3} \sigma^\mu _R
\left(\begin{array}{c} \tilde{\nu} \\
-\tilde{e}\\ \tilde{u}\\ -\tilde{d} \end{array} \right)_R
\hspace*{-0.4cm}+g_2 \omega_{\mu1} \sigma^\mu _R
\left(\begin{array}{c} 0 \\
0\\ \tilde{u}\\ \tilde{d} \end{array} \right)_R
\hspace*{-0.4cm}+i g_2 \omega_{\mu2} \sigma^\mu _R
\left(\begin{array}{c} 0 \\
0\\ \tilde{u}\\ \tilde{d} \end{array} \right)_R
\right)+\frac{1}{4}{\Omega_i}^i
\left(\begin{array}{c} \tilde{\nu} \\
\tilde{e}\\ \tilde{u}\\ \tilde{d} \end{array} \right)_R
\end{array}$$
and
$$\begin{array}{l}
0= i \sigma^\mu_L  \partial_ \mu \left(\begin{array}{c} \tilde{\nu} \\
\tilde{e}\\ \tilde{u}\\ \tilde{d} \end{array} \right)_L
\hspace*{-0.4cm}-k\left(\begin{array}{c} -l\tilde{\nu} \\
l\tilde{e}\\ q\tilde{u}\\ q\tilde{d} \end{array} \right)_{R}
\hspace*{-0.4cm}-\bar{k}\frac{v^2}{l}\left(\begin{array}{c}
\tilde{\nu} \\
0\\ 0\\ 0 \end{array} \right)_{R}
\hspace*{-0.4cm}-\frac{1}{2} \left( g_1 \omega_{\mu0}\sigma^\mu_L
\left(\begin{array}{c} \tilde{\nu} \\
\tilde{e}\\ \tilde{u}\\ \tilde{d} \end{array}
\right)_L\right.\\[0.4cm]\left.
+g_2 \omega_{\mu3} \sigma^\mu _L
\left(\begin{array}{c} \tilde{\nu} \\
-\tilde{e}\\ \tilde{u}\\ -\tilde{d} \end{array} \right)_L
\hspace*{-0.4cm}+g_2 \omega_{\mu1} \sigma^\mu _L
\left(\begin{array}{c} \tilde{\nu} \\
\tilde{e}\\ \tilde{u}\\ \tilde{d} \end{array} \right)_L
\hspace*{-0.4cm}+i g_2 \omega_{\mu2} \sigma^\mu _L
\left(\begin{array}{c} \tilde{\nu} \\
\tilde{e}\\ \tilde{u}\\ \tilde{d} \end{array} \right)_L \right)
-\frac{1}{4}{\Omega_i}^i
\left(\begin{array}{c} \tilde{\nu} \\
\tilde{e}\\ \tilde{u}\\ \tilde{d} \end{array} \right)_L
\end{array}
$$
were $\Omega_i^i$ is the trace of the spin-gauge bosons $\Omega_{\mu
a}=\omega_{\mu 3+a}$ with $i=1..3$.
Obviously for the masses of the fermions follows
$$ m_ \nu  =  \frac{\, \tilde k v ^2}{l}- 4 kl \stackrel{!}{=} 0,
\:m_e  =  4kl , \quad m_u = m_d = 4 kq
$$
After implementing the Weinberg mixture, which is the same as in
the standard model, one gets the Dirac equations which differ from  
the standard model in following points:
\begin{itemize}
\item There exists a right-handed neutrino, but it couples to the
Z-boson only (which may have measurable consequences for the boson's  
lifetime).
\item All fermions couple to the spin-gauge bosons. Since these
are Planck massive (see below) this plays no role in the low energy limit.
\item The coupling constants of the right- and left-handed quarks
can be influenced separately by the choice of ground-state, but a
"complete" asymmetry as for the neutrino would be linked to zero or  
negative mass.
\end{itemize}
\subsection*{Boson Masses}
The mass-square matrix $M^{2 \mu }{}_\nu {}^{ij}$ for the gauge bosons is
$$
\hspace*{-0cm}
\hat{g}_ {(i)(j)}\left(4 \mbox{tr}
\{  \left[
\tau ^{(i}, \stackrel{\circ }{\tilde{\sigma}}^\rho _ L \right]\left[  
\tau ^{j)}, \stackrel{\circ}{\tilde{\sigma}}_ {\lambda R}\right]
+(L\leftrightarrow R)\}
(\delta^\mu_\rho \delta^\lambda_\nu
- \frac{1}{2}\delta^\mu_\nu \delta^\lambda_\rho)
+ \delta ^\mu_\nu \stackrel{\circ}{\phi} ^{\dagger}
\left\{  \tau ^i , \tau^j \right\} \stackrel{\circ}{\phi} \right)
$$
with $\hat{g}_ {(i)(j)}=2 \pi g_ {(i)}g_ {(j)}$.
This leads to Planck massive spin-gauge bosons \cite{prev}.
 Unfortunately it seems
impossible to generate the Z-boson mass with the tensorial Higgs
fields, so that all masses of the electroweak gauge bosons have to  
be generated with help of the scalar Higgs field.
\subsection{Tensor-Field Excitations}
To investigate the structure of field equations of the tensor-field  
we neglect the gauge-bosons. The first order field equations for  
the excitations
$\epsilon_ \mu^{\nu a}$ of the tensor field are:
\begin{eqnarray*}
\nonumber\lefteqn{\partial_ \alpha \partial^\alpha
\epsilon_{R}^{\mu\nu a}-2 \partial_ \alpha \partial^ \mu
\epsilon_{R}^{\alpha \nu a}- \frac{\mu^2}{4q^2}\eta^{\mu\nu}
(q\delta^{a 4}-l \delta^{a 3})(\epsilon_{L \alpha}^{\alpha
(-l3+q4)}+\epsilon_{R \alpha}^{\alpha (l0+q4)}) =}\\
&&\frac{i}{8}(\psi_{L}^\dag N^a\sigma_{L}^\nu \partial^ \mu
\psi_{L} - h.c.) - \frac{k}{4}(\psi_{L}^\dag N^a\sigma_{L}^\nu
\sigma_{R}^{\mu} \stackrel{(0) \hfill}{N_{R}}\psi_{R}+h.c.)
\end{eqnarray*}
with the left-handed equation respectively.
Here ${\epsilon_ \mu}^{\nu (ax+by)}$ means $a{\epsilon_ \mu}^{\nu
x}+b {\epsilon_ \mu}^{\nu y}$ and the source is developed to 0th order  
only. These equations can be divided in two classes.

The first class consists of excitations with isospin-index $a
\epsilon\{0,3,4,7\}$ and is a gravitation like interaction (see  
also \cite{prev}), were
each kind of fermions generates its own gravitation. With the
redefined isospinors
(\ref{e4})
and the 0th order
energy-momentum tensor the equation for the "quark-gravity" is
\begin{eqnarray*}
\nonumber\lefteqn{\partial_ \alpha \partial^\alpha
\epsilon_{L}^{\mu \nu (4+7) }-2 \partial_ \alpha \partial^ \mu
\epsilon_{L}^{\:\alpha \nu (4+7)}- \frac{\mu^2}{4q}\eta^{\mu \nu}
(\epsilon_{L \alpha}^{\alpha (-l3+q4)}+\epsilon_{R \alpha}^{\alpha  
(l0+q4)}) =}\\
&&\frac{1}{2q}\left(T^{\mu\nu} (\tilde u_R)-\frac{\eta^{\mu
\nu}}{2}{T_ \alpha}^\alpha(\tilde u_R)+\frac{m_e}{4}\left(\tilde u_L  
^{\dagger} \sigma_L^{[\nu}\sigma_{R}^{\mu]} \tilde u_R
+h.c.\right)\right)
\end{eqnarray*}
We get similar equations for other combinations of the excitations  
${\epsilon_ \mu}^{\nu a}$ with $a \epsilon \{0,3,4,7\}$ with the
other fermions as source. Herein the neutrino is making an exception  
since its (zero) mass is partially generated by the scalar Higgs
field, i.e.
it provides in the $(0+3)$ equation an additional source term for
the gravitation-like interaction
$$
\frac{\eta^{\mu\nu}}{2} m_e \left(\tilde \nu_L ^{\dagger}
\tilde \nu_R+\tilde \nu_R ^{\dagger} \tilde
\nu_L\right)
$$
thus violating the
equivalence-principle. However, this term cancels out in the classical  
limit.

The second class consists of excitations with isospin-index $a
\epsilon\{1,2,5,6\}$. These fields carry electrical charge. The
source terms of these equations have the form of energy-momentum
tensors but they contain the fermions in a mixed form:
\begin{eqnarray*}
\nonumber\lefteqn{\partial_ \alpha \partial^\alpha {\epsilon_{L
\mu}}^{\:\nu (1+i2) }-2 \partial_ \alpha \partial_ \mu
\epsilon_{L}^{\:\alpha \nu (1+i2)} =
\frac{-i}{4l}\left(\left(\tilde \nu_R ^{\dagger}\sigma_R^\nu
\partial_\mu \tilde e_R - h.c.
\right)\right.}
\\&&\hspace{2cm}-2kl\left(\tilde \nu_L ^{\dagger}\sigma_{L \mu}  
\sigma_R^\nu
\tilde e_R-\tilde \nu_R ^{\dagger}\sigma_R^\nu \sigma_{L \mu}\tilde  
e_L \right)\left.\right)
\end{eqnarray*}
In the equations of the first class we see that quarks and leptons  
couple with different coupling-constants to gravity.
Moreover, the fact that each kind of fermions produces a different  
gravity and couples to its own gravity only (as can be seen by
investigating the Dirac equation for the exited Higgs field) is a
contradiction to experiment. Due to the different distribution of u-  
and d-quarks in the earth, this would cause a measurable violation  
of the equivalence principle of the order of $\kappa \simeq 10^{-6}$  
in the E\"otv\"os-type experiments (for the same materials even the  
"famous" result of Fischbach \cite{Fisch} is 2 orders of magnitude  
lower).

\subsection{Transition to a Uniform Gravitational Field}
By putting certain constraints on the Higgs field it is possible to  
construct one uniform gravity for all fermions. We want,
that in the Dirac equation all fermions couple to the same
Higgs-field exited by the energy-momentum tensors of all
fermions and that all other excitations can be neglected. This can  
be done by constraining the excitations to be multiples of the
groundstate.

We then get a Dirac-equation and one Higgs field equation that have  
the same form as in the iso-scalar case. It is not clear if the
classical limit applies to neutrinos also, since they are chirally  
asymmetric. If we neglect the effects of this asymmetry in the fermionic
 energy momentum the classical
limit is exactly analog to the iso-scalar case \cite{prev}.

Unfortunately we have not yet been able to find an appropriate
Lagrange density to realize these constraints.

\section{Conclusions}
This work has shown how a tensor-type Higgs field can be
interpreted as a gravity transmitting field on a flat background
spacetime. In the 2nd order appears a discrepancy to Riemannian
gravity (but not in the case of grav. waves) which needs
to be interpreted. One way may be to include torsion \cite{Han}.

The isospinorial extension allowed to generate the chiral asymmetry
as well. But an additional scalar Higgs field had to be introduced, 
and new predictions like a possible coupling of a right handed
neutrino to Z bosons need to be investigated. Beyond this only the  
excitations
proportional to the groundstate seem relevant for gravity.

Whether the additional scalar Higgs field is really essential and how
the excitations might, if at all possible, be properly constrained is 
left to future research, as well as attempts of quantization of  
this theory.


\begin{thebibliography}{99}
\bibitem{prev}H. Dehnen, E. Hitzer, Int. J. Th. Phys. 33, p. 575 (1994);
34, p. 1981 (1995): gr-qc/9412052. E. Hitzer, The Higgs-Field Theoretic Extension of
The Spin-Gauge Theory of Gravity (Thesis), Hartung-Gorre, Konstanz, 
(1996), ISBN 3-89649-012-5
\bibitem{prev2}D. Ketterer, Diplomarbeit, Univ. Konstanz (1995);
J. Fibich, Diplomarbeit, Univ. Konstanz (1996)
\bibitem{Han}M. Hanauske, Nichtlin. Erw. d. Spin-Eichth.
der Grav. (Diplomarbeit), Univ. Konstanz (1997): http://kaluza.physik.uni-konstanz.de/DE/MH/
\bibitem{DFC}DeFelice, Clarke, Rel. on Curved Manifolds, CUP, (1990)
\bibitem{MTW}Misner, Thorne, Wheeler, Gravi\-tation. Freeman and Co. (1973)
\bibitem{Wu}C.S. Wu, et al.,
Phys.Rev. 105, 1413
\bibitem{Fisch}E. Fischbach, Phys. Rev. Letters 56, 3 (1986)
\end{thebibliography}
\end{document}